\documentclass[aps,twocolumn,showpacs,preprintnumbers,amsmath,amssymb,
nofootinbib,showkeys,floatfix]{revtex4-1}

\usepackage[english]{babel}
\usepackage[utf8x]{inputenc}

\usepackage{epsfig}
\usepackage{graphicx}
\usepackage{dcolumn}
\usepackage{latexsym}
\usepackage{graphicx}
\usepackage{enumerate}
\usepackage{float}
\usepackage{placeins}
\usepackage{supertabular}

\usepackage{hyperref} 
\hypersetup{
   colorlinks=true,
   linkcolor=blue,
   citecolor=blue,
   urlcolor=blue
}

\def\tstrut{\vrule height2.2ex depth0pt width0pt} 

\bibpunct{[}{]}{,}{n}{,}{,} 

\begin{document}

\title{The role of spin-flipping terms in hadronic transitions of 
$\mathbf{\Upsilon(4S)}$}
\author{Jorge Segovia}\email{segonza@usal.es}
\author{David R. Entem}\email{entem@usal.es}
\author{Francisco Fern\'andez}\email{fdz@usal.es}
\affiliation{Grupo de F\'isica Nuclear and Instituto Universitario de
F\'isica Fundamental y Matem\'aticas (IUFFyM) \\ Universidad de Salamanca,
E-37008 Salamanca, Spain}
\date{\today}

\begin{abstract}
Recent experimental data on the $\Upsilon(4S)\to\Upsilon(1S)\eta$ and 
$\Upsilon(4S)\to h_{b}(1P)\eta$ processes seem to contradict the naive 
expectation that hadronic transitions with spin-flipping terms should be 
suppressed with respect those without spin-flip. We analyze these transitions 
using the QCD Multipole Expansion (QCDME) approach and within a constituent 
quark model framework that has been applied successfully to the heavy-quark 
sectors during the last years. The QCDME formalism requires the computation of 
hybrid intermediate states which has been performed in a natural, 
parameter-free extension of our constituent quark model based on the Quark 
Confining String (QCS) scheme. We show that i) the M1-M1 contribution in the 
decay rate of the $\Upsilon(4S)\to\Upsilon(1S)\eta$ is important and its 
suppression until now is not justified; ii) the role played by the $L=0$ hybrid 
states, which enter in the calculation of the M1-M1 contribution, explains the 
observed enhancement in the $\Upsilon(4S)\to\Upsilon(1S)\eta$ decay width; and 
iii) the anomalously large decay rate of the $\Upsilon(4S)\to h_{b}(1P)\eta$ 
transition has the same physical origin.
\end{abstract}


\maketitle

\section{INTRODUCTION}
\label{sec:introduction}

The general way of referring to an hadronic transition 
is~\cite{Brambilla:2010cs} 
\begin{equation}
\Phi_{I} \to \Phi_{F}+h,
\label{eq:scheme}
\end{equation}
where $\Phi_{I}$ and $\Phi_{F}$ stand, respectively, for the initial and final 
states of heavy quarkonium. The emitted light hadron(s), $h$, are kinematically 
dominated by single particle ($\pi^{0}$, $\eta$, $\omega$, $\ldots$) or two 
particle ($2\pi$, $2K$, $\ldots$) states. 

Hadronic transitions are important decay modes for low-lying heavy quarkonium 
states. For instance, the first observed hadronic transition $\psi(2S)\to 
J/\psi\pi\pi $~\cite{Abrams:1975zp} has a branching fraction reported by the
Particle Data Group (PDG) of $(52.58\pm0.43)\%$~\cite{Agashe:2014kda}. 
Moreover, during the last years, hadronic transitions between heavy quarkonia 
have led to a remarkable series of discoveries helping either to establish new 
conventional heavy quarkonium states, like the $h_{b}(1P)$ and $h_{b}(2P)$ 
observed in the two-pion decay of the $\Upsilon(5S)$~\cite{Adachi:2011ji} 
state, or to extract relevant information of the so-called ``XYZ'' states, like 
in the cases of 
$X(3872)$~\cite{Aubert:2004fc,Aubert:2008gu,delAmoSanchez:2010jr} and 
$X(4260)$~\cite{Lees:2012cn}. 

The BaBar Collaboration has presented a systematic study of hadronic 
transitions between $\Upsilon(mS)$ $(m=4,3,2)$ and $\Upsilon(nS)$ $(n=2,1)$ 
states, reporting the first observation of the $\Upsilon(4S)\to 
\Upsilon(1S)\eta$~\cite{Aubert:2008az}. The measured branching fraction for 
this decay is
\begin{equation}
{\cal B}(\Upsilon(4S)\to \Upsilon(1S)\eta) = (1.96\pm0.06\pm0.09)
\times 10^{-4},
\end{equation}
which is puzzling larger than the branching fraction for $\Upsilon(4S)\to 
\Upsilon(1S)\pi^{+}\pi^{-}$, $(0.800\pm0.064\pm0.027)\times10^{-4}$, with a 
ratio between them of
\begin{equation}
R_{\eta}[\Upsilon(4S)] = \frac{\Gamma(\Upsilon(4S)\to 
\Upsilon(1S)\eta)}{\Gamma(\Upsilon(4S)\to \Upsilon(1S)\pi^{+}\pi^{-})} = 
2.41\pm0.40\pm0.12.
\end{equation}

The comparison of the $\Upsilon(mS)\to \Upsilon(nS)\eta$ transitions, which are 
spin-flipping, and the corresponding $\pi^{+}\pi^{-}$ ones, that do no require 
the $b$-quark spin to flip, is particularly interesting because we naively 
expect a suppression of the transitions with spin-flipping terms. This is 
observed for the two vector bottomonium states that are below open $b$-flavored 
threshold~\cite{Agashe:2014kda}:
\begin{equation}
\begin{split}
R_{\eta}[\Upsilon(2S)] &= (1.63\pm0.23)\times10^{-3}, \\
R_{\eta}[\Upsilon(3S)] &< 2.29\times10^{-3},
\end{split}
\end{equation}
but also in the charmonium sector with $R_{\eta}[\psi(2S)] = 
(9.75\pm0.17)\times10^{-2}$~\cite{Agashe:2014kda}. Note that even for the state 
which is just close above the $c$-flavored threshold, and thus it should 
present a similar role in the charmonium sector than the $\Upsilon(4S)$ in the 
bottomonium one, the PDG reports a value of $R_{\eta}[\psi(3770)] = 
0.47\pm0.22$.

Further insight into the anomalously large $\Upsilon(4S)\to \Upsilon(1S)\eta$ 
decay rate can be gained by searching for the transition $\Upsilon(4S)\to 
h_{b}(1P)\eta$ because it is, in principle, dominated by similar spin-flipping 
contributions. The Belle Collaboration has very recently measured for the first 
time the branching fraction ${\cal B}(\Upsilon(4S)\to h_{b}(1P)\eta) = 
(2.18\pm0.11\pm0.18)\times10^{-3}$~\cite{Tamponi:2015cca}. This branching 
fraction provides a partial decay width of $(44.69\pm6.95)\,{\rm keV}$ when 
combined with the total decay rate reported by PDG~\cite{Agashe:2014kda}. This 
value is again unexpectedly large.

The anomalous hadronic decay widths can be due to several mechanisms: 
Contribution of hadron loops~\cite{Meng:2007tk,Guo:2009wr,Guo:2010zk}; 
four-quark components in the quarkonium wave functions~\cite{Ali:2010pq}; 
internal loop radiation~\cite{DiGiacomo:2000va,Simonov:2008sw}; or, as we have 
pointed out in Ref.~\cite{Segovia:2014mca} and we will see herein, the 
existence of hybrid mesons with a mass near the one of the decaying 
resonance\footnote{A similar observation has been also done in 
Ref.~\cite{Ke:2007ih}.}.

The standard theoretical approach to study hadronic transitions is QCD 
Multipole Expansion (QCDME)~\cite{Gottfried:1977gp,Bhanot:1979af,Peskin:1979va,
Bhanot:1979vb,Voloshin:1978hc,Voloshin:1980zf}. Tung-Mow Yan was the first one 
to present a gauge-invariant formulation within this 
framework~\cite{Yan:1980uh}. Many details about QCDME in the context of the 
Kuang-Yan model can be found, for instance, in 
Refs.~\cite{Kuang:1981se,Kuang:1990kd}. The interested reader is also referred 
to the recent review~\cite{Kuang:2006me} of Yu-Ping Kuang.

This approach describes the hadronic transition as a two-step process in which 
the heavy quark system initially emits, at least, two gluons that subsequently 
hadronize into light hadrons. After the emission of the first gluon and before 
the emission of the second one, there exists a propagating intermediate state 
where the $Q\bar{Q}$ pair together with the gluon forms a hybrid state. 

Other possibility is to work in a local approximation~\cite{Voloshin:1980zf} 
which does not require the description of hybrid states but it is strictly valid 
only in the limit of infinite heavy quark mass. 

The width of a hadronic transition in QCDME depends critically on the position 
in the spectrum of the hybrid states, therefore it is important to describe 
consistently the heavy quarkonia and the hybrids using as few parameters as 
possible. A description of hybrid mesons is difficult from first principles of 
QCD~\cite{Juge:1999ie,Dudek:2008sz} and one is generally forced to use models: 
the flux-tube model~\cite{Isgur:1984bm,Barnes:1995hc}, constituent 
gluons~\cite{Horn:1977rq}, Coulomb gauge QCD~\cite{Guo:2008yz}, quark confining 
string model 
(QCS)~\cite{Tye:1975fz,Giles:1977mp,Buchmuller:1979gy,Buchmuller:1980su} or QCD 
string model~\cite{Kalashnikova:2008qr}.

The hadronization vertex is independent of the properties of heavy quarkonium 
and hybrid states. Usually, this kind of vertices are calculated using low 
energy theorems~\cite{Brown:1975dz}. This procedure limits the predictive power 
of the approach since for each new configuration of initial and final bound 
states one needs new adjustable parameters. Moreover, for the interesting case 
of the hadronic transitions involving the $\eta$ meson one must discard certain 
operators involving the gluon fields. This last point is relevant for the 
discussion of this work. 

One important feature of the QCDME approach is that the hadronic transitions 
are classified in a series of chromoelectric and chromomagnetic multipoles. The 
decay rate of the $\Upsilon(4S)\to \Upsilon(1S)\eta$ process has two leading 
multipole gluon emission terms: M1-M1 and E1-M2, whereas the leading order term 
in the $\Upsilon(4S)\to \Upsilon(1S)\pi\pi$ transition is an E1-E1 
contribution. The M1-M1 term has been usually neglected. However, the hybrid 
mesons involved in the calculation of the M1-M1 amplitude are different than 
those involved in the E1-M2/E1-E1 contributions. Therefore, we consider worthy 
to estimate in the $\Upsilon(4S)\to \Upsilon(1S)\eta$ process the M1-M1 term 
and thus the effect produced in the amplitude by the hybrid meson spectrum 
before resorting to more sophisticated or exotic mechanisms.

To do this, we will use the QCDME within the same theoretical formalism 
presented in Ref.~\cite{Segovia:2014mca}. This formalism has explained 
successfully some puzzles of the two-pion hadronic transitions in the 
charmonium sector. The branching ratios $R_{\eta}[\Upsilon(nS)]$ with 
$n=2,\,3,\,4$ will be presented herein\footnote{A discussion about the goodness 
of the approach applied to the calculation of the hadronic decays treated 
herein can be found in Appendix~\ref{app:goodness}.}. An important feature is 
that our model for hybrid mesons~\cite{Segovia:2014mca} is a natural, 
parameter-free extension of a constituent quark model 
(CQM)~\cite{Vijande:2004he} (for reviews on the CQM, see 
Refs.~\cite{Valcarce:2005em,Segovia:2013wma}) that describes quite well hadron 
phenomenology and hadronic 
reactions~\cite{Fernandez:1992xs,Garcilazo:2001md,Vijande:2004at}. Furthermore, 
the CQM has been recently applied to mesons containing heavy quarks with a 
remarkable success, describing a wide range of physical observables which 
concern spectrum~\cite{Segovia:2008zz,Segovia:2009zz,Segovia:2015dia}, strong 
reactions~\cite{Segovia:2011zza,Segovia:2012cd,Segovia:2013kg} and weak 
decays~\cite{Segovia:2011dg,Segovia:2012yh,Segovia:2013sxa}.

This manuscript is arranged as follows. In Sec.~2 we introduce the constituent 
quark model pointing out only those features which are relevant for this work. 
Section~3 is dedicated to explain the parameter-free extension of our quark 
model to describe hybrid mesons. Section~4 shows the QCDME formulation of the 
hadronic transitions highlighting its most relevant features. Our results are 
provided in Sec.~5. We finish giving some conclusions in Sec.~6.


\section{CONSTITUENT QUARK MODEL}
\label{sec:CQM}

The description of the meson spectra is based on the constituent quark model 
(CQM) proposed by Vijande {\it et al.} in Ref.~\cite{Vijande:2004he}. One must 
solve the Schr\"odinger equation with a quark-antiquark potential whose main 
pieces are: i) the Goldstone-boson exchanges between dressed constituent quarks 
which is a consequence of the dynamical chiral symmetry breaking of QCD, ii) 
the perturbative one-gluon fluctuations around the instanton vacuum, and iii) a 
phenomenological confining potential which reflects the empirical fact that 
quarks and gluons have never seen as isolated particles. Note that in the 
heavy quark sector chiral symmetry is explicitly broken and thus 
Goldstone-boson exchanges do not appear.

Further details about the CQM and the fine-tuned model parameters can be found 
in Refs.~\cite{Vijande:2004he,Segovia:2008zza,Segovia:2008zz}. Here we want to 
explain in more detail our confinement potential because its screened linear 
shape is a particular feature of the model. It is well known that multigluon 
exchanges produce an attractive linearly rising potential proportional to the 
distance between infinite heavy quarks. However, sea quarks are also important 
ingredients of the strong interaction dynamics that contribute to the screening 
of the rising potential at low momenta and eventually to the breaking of the 
quark-antiquark binding string~\cite{Bali:2005fu}. Our model try to mimic this 
behavior using the following expression
\begin{equation}
V_{\rm CON}(\vec{r}\,)=\left[-a_{c}(1-e^{-\mu_{c}r})+\Delta \right] 
(\vec{\lambda}_{q}^{c}\cdot\vec{\lambda}_{\bar{q}}^{c}),
\label{eq:conf}
\end{equation}
where $a_{c}$ and $\mu_{c}$ are parameters, $r$ is the interquark distance and 
$\vec{\lambda}_{q(\bar{q})}^{c}$ are $SU(3)$ color matrices. At short distances 
this potential presents a linear behavior with an effective confinement 
strength $\sigma=-a_{c}\,\mu_{c}\,(\vec{\lambda}^{c}_{q}\cdot 
\vec{\lambda}^{c}_{\bar{q}})$, while it becomes constant at large distances 
showing a threshold defined by $V_{\rm thr} = [-a_{c}+\Delta] 
(\vec{\lambda}^{c}_{q}\cdot \vec{\lambda}^{c}_{\bar{q}})$. No $q\bar{q}$ bound 
states can be found for energies higher than this threshold.

The screened linear potential is a key feature to reproduce the degeneracy 
pattern observed for the higher excited states of light 
mesons~\cite{Segovia:2008zza}. As we assume that confining interaction is 
flavor independent, this affects also to the different quark sectors and, in 
particular, the bottomonium spectrum. Table~\ref{tab:bbstates} shows the 
$S$-wave vector bottomonium states up to $n=4$ predicted by our quark model. We 
compare masses with the standard Cornell 
model~\cite{Eichten:1978tg,Eichten:1979ms} and the experimental data reported by 
PDG~\cite{Agashe:2014kda}. Both models reproduce quite well the experimental 
data. In the Cornell model the ground state is fitted to the experimental 
figure while in our case the model parameters are fitted to all meson sectors. 
One can see that our quark model is able to reproduce in better agreement the 
excited states which is a particular feature of the screening of the linear 
confinement potential.

\begin{table}[!t]
\begin{center}
\begin{tabular}{cccc}
\hline
\hline
\tstrut
State & Mass (CQM) & Mass (Cornell) & Mass (Exp.) \\
& (MeV) & (MeV) & (MeV) \\
\hline
\tstrut
$\Upsilon(1S)$ & $9502$  & $9460$  & $9460.30\pm0.26$ \\ 
$\Upsilon(2S)$ & $10015$ & $10050$ & $10023.26\pm0.31$ \\ 
$\Upsilon(3S)$ & $10349$ & $10400$ & $10355.2\pm0.5$ \\
$\Upsilon(4S)$ & $10607$ & $10670$ & $10579.4\pm1.2$ \\ 
\hline
\hline
\end{tabular}
\caption{\label{tab:bbstates} Masses, in MeV, of the $S$-wave vector 
bottomonium states up to $n=4$ predicted by our CQM and by the Cornell 
model~\cite{Eichten:1978tg,Eichten:1979ms}. Experimental masses are 
taken from PDG~\cite{Agashe:2014kda}. The mass of the $\Upsilon(1S)$ has been 
fitted in the Cornell potential.}
\end{center}
\end{table}


\section{A MODEL FOR HYBRIDS}
\label{sec:hybrids}

From the generic properties of QCD, we might expect to have states in which the
gluonic field itself is excited and carries $J^{PC}$ quantum numbers. A 
bound-state is called glueball when any valence quark content is absent, the 
addition of a constituent quark-antiquark pair to an excited gluonic field 
gives rise to what is called an hybrid meson. The gluonic quantum numbers 
couple to those of the $q\bar{q}$ pair. This coupling may give rise to 
so-called exotic $J^{PC}$ mesons, but also can produce hybrid mesons with 
natural quantum numbers. We are interested on describing the last ones because 
they are involved in the calculation of hadronic transitions within the QCDME 
approach.

Ab-initio QCD calculations of the hybrid (even conventional) bottomonium states 
are particularly difficult because the large mass of the $b$-quark. Therefore, 
the only way up to now to describe hybrid mesons in the bottomonium sector is 
through models. An extension of the quark model described above to include 
hybrid states has been presented in Ref.~\cite{Segovia:2014mca}. This extension 
is inspired on the Buchmuller-Tye quark-confining string (QCS) 
model~\cite{Tye:1975fz,Giles:1977mp,Buchmuller:1979gy,Buchmuller:1980su} which 
assumes that the meson is composed of a quark and antiquark linked by an 
appropriate color electric flux line: the string. Gluon excitation effects are 
described by the vibration of the string. These vibrational modes provide new 
states beyond the naive meson picture and are interpreted as hybrid mesons.

The coupled equations that describe the dynamics of the string, quark and 
antiquark sectors are highly nonlinear so that there is no hope of solving them 
completely. Then, to introduce the vibrational modes, we use the following 
approximation scheme. First, we solve the string Hamiltonian via the 
Bohr-Oppenheimer method to obtain the vibrational energies as a function of the 
interquark distance~\cite{Giles:1977mp}
\begin{equation}
\begin{split}
V_{n}(r) = \sigma(r) r\left\lbrace 1 + \frac{2n\pi}{\sigma(r)
\left[(r-2d)^{2}+4d^{2}\right]} \right\rbrace^{1/2}.
\end{split}
\end{equation}
Note that $n=0$ gives $V_{0}(r) = \sigma(r) r$ where $\sigma(r) = 
(16/3)\,a_{c}\,\left[(1-e^{-\mu_{c}r})/r\right]$ attending to 
Eq.~(\ref{eq:conf}). The parameter $d$ is the correction due to the finite heavy
quark mass 
\begin{equation}
d(m_{Q},r,\sigma,n)=\frac{\sigma r^{2}\alpha_{n}}{4(2m_{Q} + \sigma
r\alpha_{n})},
\end{equation}
where $\alpha_{n}$ is related with the shape of the vibrating string and can 
take the values $1\leq\alpha_{n}\leq\sqrt{2}$.

Second, the vibrational potential is inserted into the meson equation as an 
effective potential 
\begin{equation}
V_{\rm hyb}(r)=V_{\rm OGE}(r) + V_{\rm CON}(r) + \left[V_{n}(r) - 
\sigma(r)r\right],
\label{eq:pothyb}
\end{equation}
where $V_{\rm OGE}(r) + V_{\rm CON}(r)$ is the naive quark-antiquark potential 
in the heavy quark sector with $V_{\rm OGE}(r)=-4\alpha_{s}/3r$ and $V_{\rm 
CON}(r)$ given in Eq.~(\ref{eq:conf}). $V_{n}(r)$ is the vibrational potential 
calculated above. We must subtract the term $\sigma(r)r$ because it appears 
twice, one in $V_{\rm CON}(r)$ and the other one in $V_{n}(r)$. 

We have arrived to a description of the hybrid mesons in the heavy quark 
sector that does not include new parameters besides those of the original 
quark model\footnote{The variation of the parameter $\alpha_{n}$ within its 
range $[1,\sqrt{2}]$ modifies around $30\,{\rm MeV}$ the mass of a hybrid state 
in the bottomonium sector, we have chosen the mean value 
$\alpha_{n}=\sqrt{1.5}$.}. In that sense, our calculation of the hybrid states 
is parameter-free. Another important feature of our hybrid model is that, just 
like the naive quark model, the hybrid potential has a threshold from 
which no more bound states can be found and so we have a finite number of 
hybrid states in the spectrum.

\begin{table}[!t]
\begin{center}
 \begin{tabular}{ccc}
\hline
\hline
\tstrut
K / L & $0$ & $1$ \\
\hline
$1$  & $10571$ & $10785$ \\
$2$  & $10857$ & $10999$ \\
$3$  & $11063$ & $11175$ \\
$4$  & $11232$ & $11325$ \\
$5$  & $11374$ & $11452$ \\
$6$  & $11496$ & $11562$ \\
$7$  & $11600$ & $11657$ \\
$8$  & $11690$ & $11738$ \\
$9$  & $11766$ & $11807$ \\
$10$ & $11831$ & $11866$ \\
$11$ & $11885$ & $11913$ \\
$12$ & $11927$ & -       \\
\hline
\multicolumn{3}{c}{$V_{\rm thr} = 11943$} \\
\hline
\hline
\end{tabular}
\caption{\label{tab:bbgstates} Masses, in MeV, of those hybrid states in the 
bottomonium sector that participate in the hadronic transitions calculated in 
this paper. $K$ and $L$ are the hybrid meson quantum numbers.}
\end{center}
\end{table}

Table~\ref{tab:bbgstates} shows our theoretical prediction for those hybrid 
states in the bottomonium sector that will participate in the hadronic 
transitions in which we are interested. They are classified by the 
angular momentum $L$ and radial excitation $K$. It is important to realize here 
that we predict a hybrid meson with quantum numbers 
$\left|KL\right\rangle=\left|10\right\rangle$ which is very close in mass, 
$10571\,{\rm MeV}$, to the $\Upsilon(4S)$ state, $10607\,{\rm MeV}$. As we will 
see later, this feature have important consequences in the calculation of the 
$\Upsilon(4S)$ hadronic decay rates.


\section{HADRONIC DECAY RATES}
\label{sec:rates}

The Hamiltonian for a heavy $Q\bar{Q}$ system in QCDME is given 
by~\cite{Yan:1980uh}:
\begin{equation}
H_{\rm QCD}^{\rm eff} = H_{\rm QCD}^{(0)} + H_{\rm QCD}^{(1)} + H_{\rm 
QCD}^{(2)},
\label{eq:H_QCDME}
\end{equation}
where $H_{\rm QCD}^{(0)}$ represents the sum of the kinetic and potential 
energies of the heavy quarks, $H_{\rm QCD}^{(1)}$ is related with the color 
charge of the $Q\bar{Q}$ system (which is zero for color singlets), and 
$H_{\rm QCD}^{(2)}$ couples color singlets to octet $Q\bar{Q}$ states. 
Therefore, the hadronic transitions between eigenstates 
$\left|\Phi_{I}\right\rangle$ and $\left|\Phi_{F}\right\rangle$ of $H_{\rm 
QCD}^{(0)}$ are at least second order in $H_{\rm QCD}^{(2)}$ and the leading 
term is given by
\begin{equation}
\begin{split}
&
\left\langle \Phi_{F}h \right| H_{\rm QCD}^{(2)} \frac{1}{E_{I}-H_{\rm 
QCD}^{(0)}+i\partial_{0}-H_{\rm QCD}^{(1)}} H_{\rm QCD}^{(2)} 
\left|\Phi_{I}\right\rangle = \\ 
& 
=\sum_{KL} \left\langle \Phi_{F}h \right| H_{\rm QCD}^{(2)} \left| KL 
\right\rangle \frac{1}{E_{I}-E_{KL}} \left\langle KL \right| H_{\rm QCD}^{(2)} 
\left|\Phi_{I} \right\rangle,
\label{eq:HQCD2transition}
\end{split}
\end{equation}
where $\left| KL \right\rangle$ with associated energies $E_{KL}$ are 
intermediate states after the emission of the first gluon and before the 
emission of the second one. They are the hybrid mesons described in the former 
Section.

\begin{figure}[!t]
\begin{center}
\parbox[c]{0.50\textwidth}{
\centering
\includegraphics[width=0.45\textwidth,height=0.22\textheight]
{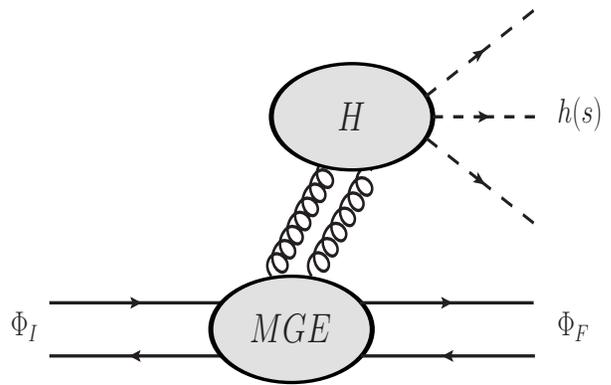}
}
\caption{\label{fig:diagram_QCDME} A hadronic transition as a two-step process: 
(1) emission of gluons from heavy quarks (MGE), and (2) the conversion of 
gluons into light hadrons (H).}
\end{center}
\end{figure}

A connection is made to the physical process in Eq.~(\ref{eq:scheme}) by 
assuming that the hadronic transition amplitude always splits into two factors 
(see Fig.~\ref{fig:diagram_QCDME}). The first one concerns the multipole gluon 
emission (MGE) from the heavy quarks and the second one is an hadronization (H) 
process describing the conversion of the emitted gluons into light hadron(s).

The MGE vertex involves the wave functions and energies of the initial and 
final quarkonium states as well as those of the intermediate hybrid mesons. All 
these quantities are calculated within our model and enter in the hadronic 
transition rate with integrals of the type\footnote{A detailed description of 
the computation of the decay rates in the single-channel approach of QCDME for 
the hadronic transitions can be found in Ref.~\cite{Kuang:1981se}. See also 
Ref.~\cite{Kuang:2006me} for an updated review and Ref.~\cite{Segovia:2014mca} 
for a calculation of two-pion spin-nonflip hadronic transitions within our 
model.}:
\begin{equation}
\begin{split}
f_{IF}^{LP_{I}P_{F}} =& \sum_{K} \frac{1}{M_{I}-M_{KL}} \left[\int dr\, 
r^{2+P_{F}} R_{F}(r)R_{KL}(r)\right] \times \\
&
\times \left[\int dr' r'^{2+P_{I}} R_{KL}(r') R_{I}(r')\right],
\label{eq:fifl}
\end{split}
\end{equation}
where $R_{I}(r)$ and $R_{F}(r)$ are, respectively, the radial wave functions of 
the initial and final states. $R_{KL}(r)$ is the radial wave function of the 
intermediate vibrational states $\left|KL\right\rangle$. The mass of the 
decaying meson is $M_{I}$, whereas the ones corresponding to the hybrid states 
are $M_{KL}$.

The H vertex is at the scale of the light hadron(s) and is independent of the 
properties of the heavy quarkonia and hybrid states. There are two ways of 
calculating the matrix elements associated with the H factor: (H1) using 
PCAC and soft pion techniques~\cite{Yan:1980uh,Brown:1975dz} or (H2) 
approximating the hadronic transition rates by 2-gluon emission 
rates~\cite{Kuang:1981se}, for instance, $\Gamma(\Phi_{I}\to \Phi_{F}\eta) 
\backsimeq \Gamma(\Phi_{I}\to \Phi_{F}(gg)_{0^{-}})$ where the two gluons are 
projected into $J^{P}=0^{-}$ to simulate the $\eta$ meson. 

Certainly, the H2-approach does not take in very detail the conversion of 
gluonic field(s) into a single $\eta$ meson. This can be view as a crude 
approximation and one should expect large uncertainties in its predictions. 
However, it is the only formalism that allows us to treat consistently the 
M1-M1 amplitude. One could estimate the uncertainties introduced by the 
H2-approach using the $\eta$ fragmentation function for the gluons at the scale 
of interest. This universal nonperturbative object has been calculated, {\it 
e.g.}, in Ref.~\cite{Aidala:2010bn} at NLO accuracy and at the scale 
$\mu=1\,{\rm GeV}$. The range of values significant for the processes studied 
here is between $0.25$ and $0.50$. Therefore, the results predicted by the 
H2-approach could have a systematic uncertainty of about $50\%$ but not orders 
of magnitude that can change drastically our conclusions.

The H1 and H2 approaches involve unknown coefficients. However, the difference 
between the two is that the H1-approach needs, at least, one coefficient for 
each multipole matrix element (we will denote these coefficients by $C_{i}$'s), 
whereas in the H2-approach all these matrix elements are written in function of 
only two parameters ($g_{E}$ and $g_{M}$).

We have mentioned in the Introduction that the leading multipoles of an $\eta$ 
transition between spin-triplet $S$-wave states are M1-M1 and E1-M2. Therefore, 
the matrix element is given schematically by
\begin{equation}
{\cal M}(^{3}S_{1}\to\,^{3}S_{1} + \eta) = {\cal M}_{M1M1} + {\cal M}_{E1M2}.
\end{equation}
The most extended version of the QCDME formalism, the H1-approach, only 
considers the E1-M2 multipole gluon emission because it is difficult to find a 
relation to fix the M1-M1 corresponding parameter, whereas the H2-approach 
considers both of them. There is no reason to neglect the M1-M1 contribution and 
this will be a key feature in order to explain the large value of the 
$\Upsilon(4S)\to \Upsilon(1S)\eta$ decay rate.

For completeness and because it is important for the discussion in the next 
Section, we show here the expressions in both approaches for the 
$^{3}S_{1}\to\,^{3}S_{1} + \pi\pi$ transition:
\begin{eqnarray}
\Gamma_{\pi\pi}^{H1} &=& G \, C_{1}^{2} \, |f_{IF}^{111}|^{2}, 
\label{eq:Gpipi1} \\
\Gamma_{\pi\pi}^{H2} &=& \left( \frac{g_{E}^{2}}{2} \right)^{2} \, 
\frac{(M_{I}-M_{F})^{7}}{1890\pi^{3}} \,  |f_{IF}^{111}|^{2},
\label{eq:Gpipi2}
\end{eqnarray}
and for the $^{3}S_{1}\to\,^{3}S_{1} + \eta$ transition:
\begin{eqnarray}
\Gamma_{\eta}^{H1} &=& \frac{8\pi^{2}}{27} \, 
\frac{M_{F}C_{3}^{2}}{M_{I}m_{Q}^{2}} \, |\vec{q}\,|^{3} \, |f_{IF}^{111}|^{2}, 
\label{eq:Geta1} \\
\Gamma_{\eta}^{H2} &=& \left(\frac{g_{M}^{2}}{3m_{Q}^{2}}\right)^{2} \,
\frac{1}{12\pi^{3}} \, \frac{(M_{I}-M_{F})^{7}}{140} \, |f_{IF}^{000}|^{2} +
\nonumber \\
&
+& \left(\frac{g_{E}g_{M}}{3m_{Q}}\right)^{2} \, \frac{1}{12\pi^{3}} \,
\frac{(M_{I}-M_{F})^{9}}{6804} \, |f_{IF}^{111}|^{2},
\label{eq:Geta2}
\end{eqnarray}
where the factor $G$ is a phase-space integral defined in Eq.~(2.4) of 
Ref.~\cite{Kuang:1981se}, $\vec{q}$ is the momentum of $\eta$ and 
$f_{IF}^{LP_{I}P_{F}}$ has been defined in Eq.~(\ref{eq:fifl}).


\section{RESULTS}
\label{sec:results}

The transitions $\Upsilon(2S)\to\Upsilon(1S)\pi^{+}\pi^{-}$ and 
$\Upsilon(2S)\to\Upsilon(1S)\eta$ help us to fix our unknown coefficients. 
The experimental data~\cite{Agashe:2014kda}
\begin{equation}
\Gamma(\Upsilon(2S)\to\Upsilon(1S)\pi^{+}\pi^{-}) = 5.71\pm0.48\,{\rm keV}, 
\end{equation}
fixes the constant $C_{1}$ in the H1-approach. Once we get $C_{1}$ the 
coefficient $g_{E}$ is determined through Eqs.~(\ref{eq:Gpipi1}) 
and~(\ref{eq:Gpipi2}). Our values of these two constants are\footnote{Note 
that the value reported here for $C_{1}$ differs to the one used in our 
previous work~\cite{Segovia:2014mca}: $9.69\times10^{-3}$, we decided to fit 
here the bottomonium case instead of the charmonium one in order to eliminate 
sources of uncertainty. One can realize that the difference is small 
$(\sim\!\!10\%)$ and it is not going to change our conclusions.}:
\begin{equation}
\begin{split}
C_{1} &= 7.69\times10^{-3}, \\
g_{E} &= 2.17,
\end{split}
\end{equation}
which compare well, for instance, with the ones used in the Kuang-Yan 
model~\cite{Kuang:1981se}.

Now, the constants $C_{3}$ and $g_{M}$ are fixed to the experimental figure 
of the $\Upsilon(2S)\to\Upsilon(1S)\eta$ decay rate~\cite{Agashe:2014kda}
\begin{equation}
\Gamma(\Upsilon(2S)\to\Upsilon(1S)\eta) = (9.27\pm1.49)\times10^{-3}\,{\rm keV},
\end{equation}
and we obtain:
\begin{equation}
\begin{split}
C_{3} &= 2.96\times10^{+6} \\
g_{M} &= 5.70
\end{split}
\end{equation}
which are also compatible with the values reported in Ref.~\cite{Kuang:1981se} 
using the Kuang-Yan model.

\begin{table*}[!t]
\begin{center}
\scalebox{0.90}{\begin{tabular}{lcccc}
\hline
\hline
\tstrut
Process & & H1-approach & H2-approach & Experiment~\cite{Agashe:2014kda} \\
\hline
\tstrut
$\Upsilon(2S)\to\Upsilon(1S)\pi^{+}\pi^{-}$ & (keV) & $5.71$ & $5.71$ & 
$5.71\pm0.48$ \\  
$\Upsilon(2S)\to\Upsilon(1S)\eta$ & (keV) & $9.27\times10^{-3}$ & 
$9.27\times10^{-3}$ & $(9.27\pm1.49)\times10^{-3}$ \\
$R_{\eta}[\Upsilon(2S)]$ & & $1.62\times10^{-3}$ & $1.62\times10^{-3}$ & 
$(1.64\pm0.25)\times10^{-3}$ \\
\hline
\tstrut
$\Upsilon(3S)\to\Upsilon(1S)\pi^{+}\pi^{-}$ & (keV) & $1.18$ & $0.80$ & 
$0.89\pm0.08$ \\
$\Upsilon(3S)\to\Upsilon(1S)\eta$ & (keV) & $7.59\times10^{-3}$ & 
$20.58\times10^{-3}$ & $<2.03\times10^{-3}$ \\ 
$R_{\eta}[\Upsilon(3S)]$ & & $6.43\times10^{-3}$ & $25.7\times10^{-3}$ & 
$<2.29\times10^{-3}$ \\
\hline
\tstrut
$\Upsilon(4S)\to\Upsilon(1S)\pi^{+}\pi^{-}$ & (keV) & $4.01$ & $2.54$ & 
$1.66\pm0.24$ \\
$\Upsilon(4S)\to\Upsilon(1S)\eta$ & (keV) & $12.60\times10^{-3}$ & $6.05$ & 
$4.02\pm0.76$ \\
$R_{\eta}[\Upsilon(4S)]$ & & $3.14\times10^{-3}$ & $2.38$ & $2.42\pm0.39$ \\
\hline
\hline
\end{tabular}}
\caption{\label{tab:pipietaWidths} The decay widths, in keV, of 
the $\Upsilon(3S)$ and $\Upsilon(4S)$ hadronic transitions into 
$\Upsilon(1S)\pi^{+}\pi^{-}$ and $\Upsilon(1S)\eta$ channels. The ratio 
$R_{\eta}[\Upsilon(nS)] = \Gamma(\Upsilon(nS)\to\Upsilon(1S)\eta) / 
\Gamma(\Upsilon(nS)\to \Upsilon(1S)\pi^{+}\pi^{-})$ is also given. The 
experimental data is taken from PDG~\cite{Agashe:2014kda}.}
\end{center}
\end{table*}

Table~\ref{tab:pipietaWidths} shows our theoretical results for the hadronic 
transitions of the $\Upsilon(3S)$ and $\Upsilon(4S)$ into 
$\Upsilon(1S)\pi^{+}\pi^{-}$ and $\Upsilon(1S)\eta$ channels. The ratio 
$R_{\eta}[\Upsilon(nS)] = \Gamma(\Upsilon(nS)\to\Upsilon(1S)\eta) / 
\Gamma(\Upsilon(nS)\to \Upsilon(1S)\pi^{+}\pi^{-})$ for $n=3,\,4$ is also 
given. The most remarkable feature of this table is the result we obtain for 
the $R_{\eta}[\Upsilon(4S)]$. While the ratio predicted within the H1-approach 
is of the order of $10^{-3}$, the calculated value using the H2-approach is of 
the order of unity and agrees nicely with the experimental 
measurement. 

Our result has a natural explanation. The rate of the $\Upsilon(4S)$ hadronic 
decay into the $\Upsilon(1S)\eta$ final state has two terms. One is a M1-M1 
contribution which involves the hybrid states of $L=0$ through the term 
$f_{IF}^{000}$ (see Eq.~(\ref{eq:Geta2})), whereas the other one is a E1-M2 
contribution which involves the $L=1$ hybrid mesons through the term 
$f_{IF}^{111}$ (see again Eq.~(\ref{eq:Geta2})). Therefore, different hybrid 
intermediate states enter in the calculation of the M1-M1 and E1-M2 terms. Our 
result indicates that the M1-M1 contribution to the hadronic decay rate can be 
(very) important because the role played by the $L=0$ hybrid spectrum.

The $L=0$ hybrid spectrum should be lower in energy than the $L=1$ spectrum. 
This is a general feature, but a model-dependent result is given in 
Table~\ref{tab:bbgstates}. The masses predicted by our model for the ground 
states of hybrid mesons with $L=0$ and $L=1$ are $10.6\,{\rm GeV}$ and 
$10.8\,{\rm GeV}$, respectively. The $\Upsilon(4S)$ is very close in mass to 
the ground state of hybrid mesons with $L=0$, and thus the mass denominator 
$M_{I}-M_{KL}$ in Eq.~(\ref{eq:fifl}) leads to a large enhancement of the 
$\Upsilon(4S)\to \Upsilon(1S)\eta$ decay rate through the M1-M1 term. The mass 
splitting between the $\Upsilon(4S)$ and the ground state of hybrid mesons with 
$L=1$ is around $0.2\,{\rm GeV}$, which is enough to predict 
$\Gamma(\Upsilon(4S)\to \Upsilon(1S)\eta)=12.92\times10^{-3}\,{\rm keV}$ when 
we consider only the E1-M2 term in the H2-approach. Note that this number 
is very close to the one obtained using the H1-approach.

Table~\ref{tab:pipietaWidths} also shows the 
$\Upsilon(3S)\to\Upsilon(1S)\pi^{+}\pi^{-}$ and 
$\Upsilon(3S)\to\Upsilon(1S)\eta$ decay rates using H1- and H2-approaches. The 
theoretical widths of the $\Upsilon(3S)\to\Upsilon(1S)\pi^{+}\pi^{-}$ process 
are compatible with the experimental data, being that of the H2-approach in 
better agreement with experiment. Only an upper limit of the decay rate for the 
$\Upsilon(3S)\to\Upsilon(1S)\eta$ process is reported by 
PDG~\cite{Agashe:2014kda}. Our predicted values are higher than this upper 
limit by a factor $\sim\!\!3$ for the H1-approach and a factor $\sim\!\!10$ for 
the H2-approach. The factor is larger in the H2-approach because there are two 
contributions to the decay rate, M1-M1 and E1-M2. Taking into account only the 
E1-M2 contribution we obtain 
$\Gamma(\Upsilon(3S)\to\Upsilon(1S)\eta)=2.59\times10^{-3}$ which is compatible 
with the experimental upper limit but still slightly higher. We encourage 
experimentalists to determine this decay rate with better precision in order to 
clarify the situation.

A couple of comments are necessary here related with the existence in the 
literature of alternative explanations to the anomalously large 
$\Upsilon(4S)\to \Upsilon(1S)\eta$ decay rate. The enhancement of the branching 
ratio $R_{\eta}[\Upsilon(4S)]$ could be attributed to neglect the effect of 
heavy-meson loops (see the related discussion for the charmonium sector in 
Refs.~\cite{Brambilla:2010cs,Guo:2010zk}). A nonrelativistic effective field 
theory (NREFT) was introduced in Ref.~\cite{Guo:2009wr} with the goal of 
determining the effect of heavy-meson loops on the hadronic transitions between 
heavy quarkonia with controlled uncertainty. The expansion parameter is the 
velocity of the heavy mesons in the intermediate state: $v_{\rm loop}$. The 
NREFT determines that a typical hadronic transition via heavy-meson loop scales 
as
\begin{equation}
v_{\rm loop}^{3} / (v_{\rm loop}^{2})^{2} \times \mbox{vertex factor},
\end{equation}
and, in the case of $R_{\eta}[\Upsilon(4S)]$, the {\it vertex factor} accounts 
for the transition between two $S$-wave bottomonia which takes place through a 
$B\bar{B}$ loop via a $P$-wave vertex. Therefore, the vertex factor is 
proportional to $v_{\rm loop}^{2}$ and the heavy-meson loop contribution scales 
as order $v_{\rm loop}$ with
\begin{equation}
v_{\rm loop} \sim \frac{\sqrt{|m[\Upsilon(4S)-2m[B]|}}{m[B]} = 
6.26\times10^{-2}.
\end{equation}
Then, heavy-meson loops cannot explain the order of magnitude measured in the 
branching ratio $R_{\eta}[\Upsilon(4S)]$.

An alternative explanation for the large decay width of the $\Upsilon(4S)\to 
\Upsilon(1S)\eta$ transition has been reported in Ref.~\cite{Simonov:2008sw}. 
Based on the Field Correlator Method (FCM)~\cite{DiGiacomo:2000va}, the authors 
of~\cite{Simonov:2008sw} assume that the $\eta$ transition between two heavy 
quarkonia proceeds via intermediate states of $BB$, $BB^{\ast}$, etc. with the 
$\eta$ meson emitted simultaneously at vertices. The dynamics of FCM is 
different than that of the QCDME and thus it can produce a different outcome 
for the decay widths. The results within FCM approach can be found in Table~3 
of Ref.~\cite{Simonov:2008sw}. The order of magnitude is reproduced for the 
$\Gamma(\Upsilon(4S)\to \Upsilon(1S)\eta)$ but a very large value, in strong 
disagreement with experiment, is also obtained for the decay rate of the 
$\Upsilon(3S)\to \Upsilon(1S)\eta$ transition. In order to alleviate this 
discrepancy, the authors modify their model parameters by $(10-15)\%$. 
Surprisingly, this keeps unmodified all partial decay widths except the 
$\Gamma(\Upsilon(3S)\to \Upsilon(1S)\eta)$ that is reduced by $3$ orders of 
magnitude due to cancellations (see again Table~3 of 
Ref.~\cite{Simonov:2008sw}).

A way to test our picture is computing other hadronic transitions of the 
$\Upsilon(4S)$ meson in which the $L=0$ hybrids are involved. One example is 
the $\Upsilon(4S)\to h_{b}(1P)\eta$ transition that has recently been observed 
by the Belle Collaboration~\cite{Tamponi:2015cca} and whose measured branching 
fraction appears to be anomalously large. The expression for the decay rate of 
the $\Upsilon(mS)\to h_{b}(nP)\eta$ hadronic transitions can be extracted from 
Eq.~(56) of Ref.~\cite{Kuang:2006me} and reads as
\begin{equation}
\begin{split}
&
\Gamma(\Upsilon(mS)\to h_{b}(nP)\eta) = \\
&
=\frac{\pi}{1144m_{Q}^{2}} \, \frac{g_{M}^{2}}{g_{E}^{2}} \, 
\frac{M_{F}|\vec{q}\,|}{M_{I}} \, 
\left(\frac{4\pi}{\sqrt{6}}f_{\pi}m_{\eta}^{2}\right)^{2} \, 
|f_{IF}^{001} + f_{IF}^{110}|^{2}.
\label{eq:spinfliphb}
\end{split}
\end{equation}

As one can see in Eq.~(\ref{eq:spinfliphb}), both $L=0$ and $L=1$ hybrid states 
are involved in this kind of transitions through the factors $f_{IF}^{001}$ and 
$f_{IF}^{110}$, respectively. We predict
\begin{equation}
\Gamma(\Upsilon(4S)\to h_{b}(1P)\eta) = 37.89\,{\rm keV},
\end{equation}
which is in good agreement with the experimental figure, $(44.69\pm6.95)\,{\rm 
keV}$. The enhancement of this decay width has the same physical origin than 
the one found in the decay rate of the $\Upsilon(4S)\to\Upsilon(1S)\eta$ 
process: we are predicting a hybrid bottomonium meson which is very close in 
mass to the $\Upsilon(4S)$. The location of the hybrid mesons in the spectrum 
is a model-dependent result but our prediction seems to explain two 
uncorrelated processes, $\Upsilon(4S)\to \Upsilon(1S)\eta$ and $\Upsilon(4S)\to 
h_{b}(1P)\eta$, even a third one considering our results in 
Ref.~\cite{Segovia:2014mca}. It is worth to remark again that the construction 
of our hybrid model does not add any new parameter of a quark model which 
explains a very large number of hadron phenomenology.


\section{CONCLUSIONS}
\label{sec:conclusions}

Recent experiments on the $\Upsilon(4S)\to\Upsilon(1S)\eta$ and 
$\Upsilon(4S)\to h_{b}(1P)\eta$ transitions have pointed out their anomalously 
large decay rates. This seems to contradict the naive expectation that hadronic 
transitions with spin-flipping terms should be suppressed with respect those 
that do not have these terms.

We have studied these transitions within the theoretical framework of a 
constituent quark model that has been applied successfully to a wide range of 
hadron observables. In particular, this model has been used in the last few 
years to study spectra, strong reactions and weak decays in the heavy quark 
sectors. Therefore, we consider that the quark model parameters are very well 
constrained for the study of hadronic transitions between heavy quarkonia.

The calculation of the hadronic decay rates has been performed using the QCDME 
approach. This formalism requires the computation of a hybrid meson spectrum. 
We have calculated the hybrid states using a natural, parameter-free extension 
of our quark model based on the Quark Confining String scheme. 

The rate of the $\Upsilon(4S)\to\Upsilon(1S)\eta$ hadronic decay has two 
leading multipole gluon emission terms: M1-M1 and E1-M2. The M1-M1 term has been 
usually neglected. We have shown that the hybrid mesons involved in the 
calculation of the M1-M1 and E1-M2 contributions are different. The M1-M1 
contribution involves the hybrid states of $L=0$, whereas the E1-M2 
contribution involves the $L=1$ hybrid mesons. The $\Upsilon(4S)$ is very close 
in mass to our predicted ground state of hybrid mesons with $L=0$, and this 
leads to a large enhancement of the $\Upsilon(4S)\to \Upsilon(1S)\eta$ decay 
rate through the M1-M1 term.

A way to test our prediction against others is looking for hadronic 
transitions that involve the $L=0$ hybrids and see if similar enhancements have 
been observed experimentally. An example is the anomalously large decay rate of 
the $\Upsilon(4S)\to h_{b}(1P)\eta$ process measured recently by the Belle 
Collaboration. We have shown here that this enhancement has the same physical 
origin than the one found in the decay $\Upsilon(4S)\to \Upsilon(1S)\eta$. We 
must admit that the location of the hybrid mesons in the spectrum is a 
model-dependent result but our prediction seems to explain the two processes 
above and even a third one considering our results in 
Ref.~\cite{Segovia:2014mca}.


\begin{acknowledgments}
This work has been partially funded by Ministerio de Ciencia y Tecnolog\'\i a
under Contract no. FPA2013-47443-C2-2-P, by the European Community-Research
Infrastructure Integrating Activity ``Study of Strongly Interacting Matter''
(HadronPhysics3 Grant no. 283286), by the Spanish Ingenio-Consolider 2010
Program CPAN (CSD2007-00042) and by the Spanish Excellence Network on Hadronic 
Physics FIS2014-57026-REDT. J. Segovia acknowledges financial support from a
postdoctoral IUFFyM contract of the Universidad de Salamanca, Spain.
\end{acknowledgments}


\appendix

\section{GOODNESS OF THE QCDME APPROACH}
\label{app:goodness}

The QCDME approach is based on performing an expansion of the gluon field 
$A^{a}_{\mu}(\vec{x},t)$ in Taylor series of $(\vec{x}-\vec{X})$ at the center 
of mass position $\vec{X}$\footnote{$\vec{X} \equiv (\vec{x}_{1}+\vec{x}_2)/2$ 
is the center of mass position of $Q$ and $\bar{Q}$, and $\vec{x}$ denotes 
$\vec{x}_{1}$ or $\vec{x}_{2}$.}:
\begin{equation}
\begin{split}
A^{a}_{0}(\vec{x},t) &= A^{a}_{0}(\vec{X},t)-(\vec{x}-\vec{X})\cdot
\vec{E}^{a}(\vec{X},t)+\ldots, \\
\vec{A}^{a}(\vec{x},t) &= -\frac{1}{2}(\vec{x}-\vec{X}) \times 
\vec{B}^{a}(\vec{X},t) + \ldots,
\end{split}
\end{equation}
where $\vec{E}^{a}$ and $\vec{B}^{a}$ are color-electric and color-magnetic
fields, respectively.

The Hamiltonian formulation is more convenient when one wants to follow a 
nonrelativistic formalism. The corresponding Hamiltonian derived from the above 
formulation is given in Eq.~(\ref{eq:H_QCDME}) where, more 
explicitly, $H_{\rm QCD}^{(2)}$ is
\begin{equation}
H_{\rm QCD}^{(2)} \equiv -\vec{d}_{a} \cdot \vec{E}^{a}(\vec{X},t) - \vec{m}_{a} 
\cdot \vec{B}^{a}(\vec{X},t) + \ldots,
\label{eq:H2}
\end{equation}
with
\begin{equation}
\begin{split}
\vec{d}_{a} &\equiv g_{E} \int (\vec{x}-\vec{X})
\Psi^{\dagger}(\vec{x},t) \frac{\lambda_{a}}{2} \Psi(\vec{x},t) d^{3}x, \\
\vec{m}_{a} &\equiv \frac{g_{M}}{2} \int (\vec{x}-\vec{X}) \times
\Psi^{\dagger}(\vec{x},t) \, \vec{\gamma} \, \frac{\lambda_{a}}{2}
\Psi(\vec{x},t) d^{3}x,
\label{eq:multipole}
\end{split}
\end{equation}
the color-electric dipole moment (E1) and the color magnetic dipole moment 
(M1) of the $Q\bar{Q}$ system, respectively. Higher-order terms (not shown) give 
rise to higher-order electric (E2, E3, ...) and magnetic moments (M2, ...).

Since $H_{\rm QCD}^{(2)}$ couples color-singlet to octet quark-antiquark 
states, the hadronic transitions are at least second order in $H_{\rm 
QCD}^{(2)}$ and the leading term is given by Eq.~(\ref{eq:HQCD2transition}). 
One realizes that the hadronic decay rate should scale as
\begin{equation}
\left\langle \Phi_{F}h \right| H_{\rm QCD}^{(2)} \left| KL \right\rangle 
\times 
\left\langle KL \right| H_{\rm QCD}^{(2)} \left|\Phi_{I} \right\rangle.
\end{equation}

For instance and without lost of generality, let us write the amplitude of the 
QCD multipole expansion approach to the spin-nonflip two pion hadronic 
transitions between heavy quarkonia:
\begin{equation}
\begin{split}
{\cal M}_{E1E1} =& i\frac{g_{E}^{2}}{6} \sum_{KL}
\frac{\left\langle\right.\!\! \Phi_{F}|x_k|KL \!\!\left.\right\rangle
\left\langle\right.\!\! KL|x_l|\Phi_I \!\!\left.\right\rangle}{E_I-E_{KL}}
 \times \\
&
\times \left\langle\right.\!\! h|E^{a}_{k} E^{a}_{l}|0 \!\!\left.\right\rangle,
\label{eq:E1E1}
\end{split}
\end{equation}
where one is assuming factorization of the heavy-quark interaction and the 
production of light hadrons. From Eq.~(\ref{eq:E1E1}), the hadronic 
decay rate is proportional to
\begin{equation}
(k_{F}a_{F}) \times (k_{I}a_{I}) = \left[k_{F} \left\langle \Phi_{F} | x_k | KL 
\right\rangle \right] \times \left[ k_{I} \left\langle  KL | x_l | \Phi_{I} 
\right\rangle \right],
\label{eq:scale}
\end{equation}
and not, as usually considered, to $(ka)^{2}$ with $a$ the average size 
of the heavy quarkonium states involved in the hadronic transition and $k$ the 
available momentum of the process.

Table~\ref{tab:good} shows the relative size of the overlap terms that appear 
in Eq.~(\ref{eq:scale}) and are involved in the $\Upsilon(4S)\to 
\Upsilon(1S)\pi^{+}\pi^{-}$ transition. We compare them with those obtained for 
the $\Upsilon(2S)$ and $\Upsilon(3S)$ states in which usually QCDME approach is 
considered to work well. One can conclude that QCDME approach works at the same 
level for the $\Upsilon(4S)$ state than for the $\Upsilon(2S)$ and 
$\Upsilon(3S)$ states.

\begin{table}
\begin{center}
\scalebox{0.70}{\begin{tabular}{cccc|ccc|ccc}
\hline
\hline
\tstrut
Initial & Final & $L$ & $K$ & $k_{F}$ & $a_{F}=\left\langle \Phi_{F} | r | KL 
\right\rangle$ & $k_{F}a_{F}$ & $k_{I}$ & $a_{I}=\left\langle KL | r | \Phi_{I} 
\right\rangle$ & $k_{I}a_{I}$ \\
& & & & (GeV) & (fm) & & (GeV) & (fm) & \\ 
\hline
\tstrut
$\Upsilon(4S)$ & $\Upsilon(1S)$ & $1$ & $1$  & $1.24$ & $0.20$   & $1.27$  & $0.21$ & $0.016$ & $0.017$ \\
               &                & $1$ & $2$  & $1.43$ & $0.10$   & $0.74$  & $0.43$ & $0.41$  & $0.89$ \\
               &                & $1$ & $3$  & $1.58$ & $0.059$  & $0.47$  & $0.61$ & $0.84$  & $2.62$ \\
               &                & $1$ & $4$  & $1.71$ & $0.038$  & $0.33$  & $0.77$ & $0.054$ & $0.21$ \\               
               &                & $1$ & $5$  & $1.82$ & $0.027$  & $0.25$  & $0.91$ & $0.094$ & $0.43$ \\
               &                & $1$ & $6$  & $1.91$ & $0.020$  & $0.19$  & $1.03$ & $0.10$  & $0.54$ \\
               &                & $1$ & $7$  & $1.99$ & $0.015$  & $0.15$  & $1.13$ & $0.087$ & $0.50$ \\
               &                & $1$ & $8$  & $2.06$ & $0.012$  & $0.13$  & $1.22$ & $0.070$ & $0.43$ \\
               &                & $1$ & $9$  & $2.11$ & $0.0098$ & $0.10$  & $1.30$ & $0.056$ & $0.37$ \\
               &                & $1$ & $10$ & $2.16$ & $0.0081$ & $0.088$ & $1.36$ & $0.045$ & $0.31$ \\
               &                & $1$ & $11$ & $2.20$ & $0.0067$ & $0.075$ & $1.42$ & $0.037$ & $0.26$ \\[2ex]
$\Upsilon(3S)$ & $\Upsilon(1S)$ & $1$ & $1$  & $1.24$ & $0.20$   & $1.27$  & $0.44$ & $0.21$  & $0.46$ \\
               &                & $1$ & $2$  & $1.43$ & $0.10$   & $0.74$  & $0.66$ & $0.67$  & $2.26$ \\
               &                & $1$ & $3$  & $1.58$ & $0.059$  & $0.47$  & $0.85$ & $0.066$ & $0.29$ \\
               &                & $1$ & $4$  & $1.71$ & $0.038$  & $0.33$  & $1.02$ & $0.13$  & $0.66$ \\               
               &                & $1$ & $5$  & $1.82$ & $0.027$  & $0.25$  & $1.16$ & $0.10$  & $0.61$ \\
               &                & $1$ & $6$  & $1.91$ & $0.020$  & $0.19$  & $1.28$ & $0.078$ & $0.51$ \\
               &                & $1$ & $7$  & $1.99$ & $0.015$  & $0.15$  & $1.38$ & $0.059$ & $0.41$ \\
               &                & $1$ & $8$  & $2.06$ & $0.012$  & $0.13$  & $1.48$ & $0.045$ & $0.34$ \\
               &                & $1$ & $9$  & $2.11$ & $0.0098$ & $0.10$  & $1.55$ & $0.035$ & $0.28$ \\
               &                & $1$ & $10$ & $2.16$ & $0.0081$ & $0.088$ & $1.62$ & $0.028$ & $0.23$ \\
               &                & $1$ & $11$ & $2.20$ & $0.0067$ & $0.075$ & $1.68$ & $0.023$ & $0.19$ \\[2ex]
$\Upsilon(2S)$ & $\Upsilon(1S)$ & $1$ & $1$  & $1.24$ & $0.20$   & $1.27$  & $0.79$ & $0.44$  & $1.76$ \\
               &                & $1$ & $2$  & $1.43$ & $0.10$   & $0.74$  & $1.02$ & $0.16$  & $0.84$ \\
               &                & $1$ & $3$  & $1.58$ & $0.059$  & $0.47$  & $1.22$ & $0.13$  & $0.82$ \\
               &                & $1$ & $4$  & $1.71$ & $0.038$  & $0.33$  & $1.39$ & $0.089$ & $0.63$ \\               
               &                & $1$ & $5$  & $1.82$ & $0.027$  & $0.25$  & $1.53$ & $0.061$ & $0.47$ \\
               &                & $1$ & $6$  & $1.91$ & $0.020$  & $0.19$  & $1.66$ & $0.044$ & $0.37$ \\
               &                & $1$ & $7$  & $1.99$ & $0.015$  & $0.15$  & $1.77$ & $0.032$ & $0.29$ \\
               &                & $1$ & $8$  & $2.06$ & $0.012$  & $0.13$  & $1.86$ & $0.025$ & $0.23$ \\
               &                & $1$ & $9$  & $2.11$ & $0.0098$ & $0.10$  & $1.94$ & $0.020$ & $0.19$ \\
               &                & $1$ & $10$ & $2.16$ & $0.0081$ & $0.088$ & $2.01$ & $0.016$ & $0.16$ \\
               &                & $1$ & $11$ & $2.20$ & $0.0067$ & $0.075$ & $2.07$ & $0.013$ & $0.14$ \\
\hline
\hline
\end{tabular}}
\caption{\label{tab:good} The relative size of the overlap terms that are 
involved in the hadronic transitions 
$\Upsilon(nS)\to\Upsilon(1S)\pi^{+}\pi^{-}$ with $n=2,\,3,\,4$.}
\end{center}
\end{table}


\section*{References}
\bibliographystyle{apsrev}
\bibliography{HadronicTransitions2}

\begin{thebibliography}{60}
\expandafter\ifx\csname natexlab\endcsname\relax\def\natexlab#1{#1}\fi
\expandafter\ifx\csname bibnamefont\endcsname\relax
  \def\bibnamefont#1{#1}\fi
\expandafter\ifx\csname bibfnamefont\endcsname\relax
  \def\bibfnamefont#1{#1}\fi
\expandafter\ifx\csname citenamefont\endcsname\relax
  \def\citenamefont#1{#1}\fi
\expandafter\ifx\csname url\endcsname\relax
  \def\url#1{\texttt{#1}}\fi
\expandafter\ifx\csname urlprefix\endcsname\relax\def\urlprefix{URL }\fi
\providecommand{\bibinfo}[2]{#2}
\providecommand{\eprint}[2][]{\url{#2}}

\bibitem[{\citenamefont{Brambilla et~al.}(2011)\citenamefont{Brambilla,
  Eidelman, Heltsley, Vogt, Bodwin et~al.}}]{Brambilla:2010cs}
\bibinfo{author}{\bibfnamefont{N.}~\bibnamefont{Brambilla}},
  \bibinfo{author}{\bibfnamefont{S.}~\bibnamefont{Eidelman}},
  \bibinfo{author}{\bibfnamefont{B.}~\bibnamefont{Heltsley}},
  \bibinfo{author}{\bibfnamefont{R.}~\bibnamefont{Vogt}},
  \bibinfo{author}{\bibfnamefont{G.}~\bibnamefont{Bodwin}},
  \bibnamefont{et~al.}, \bibinfo{journal}{Eur. Phys. J.}
  \textbf{\bibinfo{volume}{C71}}, \bibinfo{pages}{1534} (\bibinfo{year}{2011}).

\bibitem[{\citenamefont{Abrams et~al.}(1975)}]{Abrams:1975zp}
\bibinfo{author}{\bibfnamefont{G.}~\bibnamefont{Abrams}} \bibnamefont{et~al.},
  \bibinfo{journal}{Phys. Rev. Lett.} \textbf{\bibinfo{volume}{34}},
  \bibinfo{pages}{1181} (\bibinfo{year}{1975}).

\bibitem[{\citenamefont{Olive et~al.}(2014)}]{Agashe:2014kda}
\bibinfo{author}{\bibfnamefont{K.}~\bibnamefont{Olive}} \bibnamefont{et~al.}
  (\bibinfo{collaboration}{Particle Data Group}), \bibinfo{journal}{Chin.
  Phys.} \textbf{\bibinfo{volume}{C38}}, \bibinfo{pages}{090001}
  (\bibinfo{year}{2014}).

\bibitem[{\citenamefont{Adachi et~al.}(2012)}]{Adachi:2011ji}
\bibinfo{author}{\bibfnamefont{I.}~\bibnamefont{Adachi}} \bibnamefont{et~al.}
  (\bibinfo{collaboration}{Belle}), \bibinfo{journal}{Phys. Rev. Lett.}
  \textbf{\bibinfo{volume}{108}}, \bibinfo{pages}{032001}
  (\bibinfo{year}{2012}).

\bibitem[{\citenamefont{Aubert et~al.}(2004)}]{Aubert:2004fc}
\bibinfo{author}{\bibfnamefont{B.}~\bibnamefont{Aubert}} \bibnamefont{et~al.}
  (\bibinfo{collaboration}{BaBar}), \bibinfo{journal}{Phys. Rev. Lett.}
  \textbf{\bibinfo{volume}{93}}, \bibinfo{pages}{041801}
  (\bibinfo{year}{2004}).

\bibitem[{\citenamefont{Aubert et~al.}(2008{\natexlab{a}})}]{Aubert:2008gu}
\bibinfo{author}{\bibfnamefont{B.}~\bibnamefont{Aubert}} \bibnamefont{et~al.}
  (\bibinfo{collaboration}{BaBar}), \bibinfo{journal}{Phys. Rev.}
  \textbf{\bibinfo{volume}{D77}}, \bibinfo{pages}{111101}
  (\bibinfo{year}{2008}{\natexlab{a}}).

\bibitem[{\citenamefont{del Amo~Sanchez et~al.}(2010)}]{delAmoSanchez:2010jr}
\bibinfo{author}{\bibfnamefont{P.}~\bibnamefont{del Amo~Sanchez}}
  \bibnamefont{et~al.} (\bibinfo{collaboration}{BaBar}),
  \bibinfo{journal}{Phys. Rev.} \textbf{\bibinfo{volume}{D82}},
  \bibinfo{pages}{011101} (\bibinfo{year}{2010}).

\bibitem[{\citenamefont{Lees et~al.}(2012)}]{Lees:2012cn}
\bibinfo{author}{\bibfnamefont{J.}~\bibnamefont{Lees}} \bibnamefont{et~al.}
  (\bibinfo{collaboration}{BaBar}), \bibinfo{journal}{Phys. Rev.}
  \textbf{\bibinfo{volume}{D86}}, \bibinfo{pages}{051102}
  (\bibinfo{year}{2012}).

\bibitem[{\citenamefont{Aubert et~al.}(2008{\natexlab{b}})}]{Aubert:2008az}
\bibinfo{author}{\bibfnamefont{B.}~\bibnamefont{Aubert}} \bibnamefont{et~al.}
  (\bibinfo{collaboration}{BaBar}), \bibinfo{journal}{Phys. Rev.}
  \textbf{\bibinfo{volume}{D78}}, \bibinfo{pages}{112002}
  (\bibinfo{year}{2008}{\natexlab{b}}).

\bibitem[{\citenamefont{Tamponi et~al.}(2015)}]{Tamponi:2015cca}
\bibinfo{author}{\bibfnamefont{U.}~\bibnamefont{Tamponi}} \bibnamefont{et~al.}
  (\bibinfo{collaboration}{The Belle Collaboration}) (\bibinfo{year}{2015}),
  \eprint{arXiv:hep-ex/1506.08914}.

\bibitem[{\citenamefont{Meng and Chao}(2008)}]{Meng:2007tk}
\bibinfo{author}{\bibfnamefont{C.}~\bibnamefont{Meng}} \bibnamefont{and}
  \bibinfo{author}{\bibfnamefont{K.-T.} \bibnamefont{Chao}},
  \bibinfo{journal}{Phys. Rev.} \textbf{\bibinfo{volume}{D77}},
  \bibinfo{pages}{074003} (\bibinfo{year}{2008}).

\bibitem[{\citenamefont{Guo et~al.}(2009)\citenamefont{Guo, Hanhart, and
  Meissner}}]{Guo:2009wr}
\bibinfo{author}{\bibfnamefont{F.-K.} \bibnamefont{Guo}},
  \bibinfo{author}{\bibfnamefont{C.}~\bibnamefont{Hanhart}}, \bibnamefont{and}
  \bibinfo{author}{\bibfnamefont{U.-G.} \bibnamefont{Meissner}},
  \bibinfo{journal}{Phys. Rev. Lett.} \textbf{\bibinfo{volume}{103}},
  \bibinfo{pages}{082003} (\bibinfo{year}{2009}).

\bibitem[{\citenamefont{Guo et~al.}(2010)\citenamefont{Guo, Hanhart, Li,
  Meissner, and Zhao}}]{Guo:2010zk}
\bibinfo{author}{\bibfnamefont{F.-K.} \bibnamefont{Guo}},
  \bibinfo{author}{\bibfnamefont{C.}~\bibnamefont{Hanhart}},
  \bibinfo{author}{\bibfnamefont{G.}~\bibnamefont{Li}},
  \bibinfo{author}{\bibfnamefont{U.-G.} \bibnamefont{Meissner}},
  \bibnamefont{and} \bibinfo{author}{\bibfnamefont{Q.}~\bibnamefont{Zhao}},
  \bibinfo{journal}{Phys. Rev.} \textbf{\bibinfo{volume}{D82}},
  \bibinfo{pages}{034025} (\bibinfo{year}{2010}).

\bibitem[{\citenamefont{Ali et~al.}(2011)\citenamefont{Ali, Hambrock, and
  Mishima}}]{Ali:2010pq}
\bibinfo{author}{\bibfnamefont{A.}~\bibnamefont{Ali}},
  \bibinfo{author}{\bibfnamefont{C.}~\bibnamefont{Hambrock}}, \bibnamefont{and}
  \bibinfo{author}{\bibfnamefont{S.}~\bibnamefont{Mishima}},
  \bibinfo{journal}{Phys. Rev. Lett.} \textbf{\bibinfo{volume}{106}}
  (\bibinfo{year}{2011}), \eprint{1011.4856}.

\bibitem[{\citenamefont{Di~Giacomo et~al.}(2002)\citenamefont{Di~Giacomo,
  Dosch, Shevchenko, and Simonov}}]{DiGiacomo:2000va}
\bibinfo{author}{\bibfnamefont{A.}~\bibnamefont{Di~Giacomo}},
  \bibinfo{author}{\bibfnamefont{H.~G.} \bibnamefont{Dosch}},
  \bibinfo{author}{\bibfnamefont{V.~I.} \bibnamefont{Shevchenko}},
  \bibnamefont{and} \bibinfo{author}{\bibfnamefont{{\relax Yu}.~A.}
  \bibnamefont{Simonov}}, \bibinfo{journal}{Phys. Rept.}
  \textbf{\bibinfo{volume}{372}}, \bibinfo{pages}{319} (\bibinfo{year}{2002}).

\bibitem[{\citenamefont{Simonov and Veselov}(2009)}]{Simonov:2008sw}
\bibinfo{author}{\bibfnamefont{{\relax Yu}.~A.} \bibnamefont{Simonov}}
  \bibnamefont{and} \bibinfo{author}{\bibfnamefont{A.~I.}
  \bibnamefont{Veselov}}, \bibinfo{journal}{Phys. Lett.}
  \textbf{\bibinfo{volume}{B673}}, \bibinfo{pages}{211} (\bibinfo{year}{2009}).

\bibitem[{\citenamefont{Segovia
  et~al.}(2015{\natexlab{a}})\citenamefont{Segovia, Entem, and
  Fernandez}}]{Segovia:2014mca}
\bibinfo{author}{\bibfnamefont{J.}~\bibnamefont{Segovia}},
  \bibinfo{author}{\bibfnamefont{D.~R.} \bibnamefont{Entem}}, \bibnamefont{and}
  \bibinfo{author}{\bibfnamefont{F.}~\bibnamefont{Fernandez}},
  \bibinfo{journal}{Phys. Rev.} \textbf{\bibinfo{volume}{D91}},
  \bibinfo{pages}{014002} (\bibinfo{year}{2015}{\natexlab{a}}).

\bibitem[{\citenamefont{Ke et~al.}(2007)\citenamefont{Ke, Tang, Hao, and
  Li}}]{Ke:2007ih}
\bibinfo{author}{\bibfnamefont{H.-W.} \bibnamefont{Ke}},
  \bibinfo{author}{\bibfnamefont{J.}~\bibnamefont{Tang}},
  \bibinfo{author}{\bibfnamefont{X.-Q.} \bibnamefont{Hao}}, \bibnamefont{and}
  \bibinfo{author}{\bibfnamefont{X.-Q.} \bibnamefont{Li}},
  \bibinfo{journal}{Phys. Rev.} \textbf{\bibinfo{volume}{D76}},
  \bibinfo{pages}{074035} (\bibinfo{year}{2007}).

\bibitem[{\citenamefont{Gottfried}(1978)}]{Gottfried:1977gp}
\bibinfo{author}{\bibfnamefont{K.}~\bibnamefont{Gottfried}},
  \bibinfo{journal}{Phys. Rev. Lett.} \textbf{\bibinfo{volume}{40}},
  \bibinfo{pages}{598} (\bibinfo{year}{1978}).

\bibitem[{\citenamefont{Bhanot et~al.}(1979)\citenamefont{Bhanot, Fischler, and
  Rudaz}}]{Bhanot:1979af}
\bibinfo{author}{\bibfnamefont{G.}~\bibnamefont{Bhanot}},
  \bibinfo{author}{\bibfnamefont{W.}~\bibnamefont{Fischler}}, \bibnamefont{and}
  \bibinfo{author}{\bibfnamefont{S.}~\bibnamefont{Rudaz}},
  \bibinfo{journal}{Nucl. Phys.} \textbf{\bibinfo{volume}{B155}},
  \bibinfo{pages}{208} (\bibinfo{year}{1979}).

\bibitem[{\citenamefont{Peskin}(1979)}]{Peskin:1979va}
\bibinfo{author}{\bibfnamefont{M.~E.} \bibnamefont{Peskin}},
  \bibinfo{journal}{Nucl. Phys.} \textbf{\bibinfo{volume}{B156}},
  \bibinfo{pages}{365} (\bibinfo{year}{1979}).

\bibitem[{\citenamefont{Bhanot and Peskin}(1979)}]{Bhanot:1979vb}
\bibinfo{author}{\bibfnamefont{G.}~\bibnamefont{Bhanot}} \bibnamefont{and}
  \bibinfo{author}{\bibfnamefont{M.~E.} \bibnamefont{Peskin}},
  \bibinfo{journal}{Nucl. Phys.} \textbf{\bibinfo{volume}{B156}},
  \bibinfo{pages}{391} (\bibinfo{year}{1979}).

\bibitem[{\citenamefont{Voloshin}(1979)}]{Voloshin:1978hc}
\bibinfo{author}{\bibfnamefont{M.}~\bibnamefont{Voloshin}},
  \bibinfo{journal}{Nucl. Phys.} \textbf{\bibinfo{volume}{B154}},
  \bibinfo{pages}{365} (\bibinfo{year}{1979}).

\bibitem[{\citenamefont{Voloshin and Zakharov}(1980)}]{Voloshin:1980zf}
\bibinfo{author}{\bibfnamefont{M.~B.} \bibnamefont{Voloshin}} \bibnamefont{and}
  \bibinfo{author}{\bibfnamefont{V.~I.} \bibnamefont{Zakharov}},
  \bibinfo{journal}{Phys. Rev. Lett.} \textbf{\bibinfo{volume}{45}},
  \bibinfo{pages}{688} (\bibinfo{year}{1980}).

\bibitem[{\citenamefont{Yan}(1980)}]{Yan:1980uh}
\bibinfo{author}{\bibfnamefont{T.-M.} \bibnamefont{Yan}},
  \bibinfo{journal}{Phys. Rev.} \textbf{\bibinfo{volume}{D22}},
  \bibinfo{pages}{1652} (\bibinfo{year}{1980}).

\bibitem[{\citenamefont{Kuang and Yan}(1981)}]{Kuang:1981se}
\bibinfo{author}{\bibfnamefont{Y.-P.} \bibnamefont{Kuang}} \bibnamefont{and}
  \bibinfo{author}{\bibfnamefont{T.-M.} \bibnamefont{Yan}},
  \bibinfo{journal}{Phys. Rev.} \textbf{\bibinfo{volume}{D24}},
  \bibinfo{pages}{2874} (\bibinfo{year}{1981}).

\bibitem[{\citenamefont{Kuang et~al.}(1990)\citenamefont{Kuang, Yi, and
  Fu}}]{Kuang:1990kd}
\bibinfo{author}{\bibfnamefont{Y.-P.} \bibnamefont{Kuang}},
  \bibinfo{author}{\bibfnamefont{Y.-P.} \bibnamefont{Yi}}, \bibnamefont{and}
  \bibinfo{author}{\bibfnamefont{B.}~\bibnamefont{Fu}}, \bibinfo{journal}{Phys.
  Rev.} \textbf{\bibinfo{volume}{D42}}, \bibinfo{pages}{2300}
  (\bibinfo{year}{1990}).

\bibitem[{\citenamefont{Kuang}(2006)}]{Kuang:2006me}
\bibinfo{author}{\bibfnamefont{Y.-P.} \bibnamefont{Kuang}},
  \bibinfo{journal}{Front. Phys. China} \textbf{\bibinfo{volume}{1}},
  \bibinfo{pages}{19} (\bibinfo{year}{2006}).

\bibitem[{\citenamefont{Juge et~al.}(1999)\citenamefont{Juge, Kuti, and
  Morningstar}}]{Juge:1999ie}
\bibinfo{author}{\bibfnamefont{K.}~\bibnamefont{Juge}},
  \bibinfo{author}{\bibfnamefont{J.}~\bibnamefont{Kuti}}, \bibnamefont{and}
  \bibinfo{author}{\bibfnamefont{C.}~\bibnamefont{Morningstar}},
  \bibinfo{journal}{Phys. Rev. Lett.} \textbf{\bibinfo{volume}{82}},
  \bibinfo{pages}{4400} (\bibinfo{year}{1999}).

\bibitem[{\citenamefont{Dudek and Rrapaj}(2008)}]{Dudek:2008sz}
\bibinfo{author}{\bibfnamefont{J.~J.} \bibnamefont{Dudek}} \bibnamefont{and}
  \bibinfo{author}{\bibfnamefont{E.}~\bibnamefont{Rrapaj}},
  \bibinfo{journal}{Phys. Rev.} \textbf{\bibinfo{volume}{D78}},
  \bibinfo{pages}{094504} (\bibinfo{year}{2008}).

\bibitem[{\citenamefont{Isgur and Paton}(1985)}]{Isgur:1984bm}
\bibinfo{author}{\bibfnamefont{N.}~\bibnamefont{Isgur}} \bibnamefont{and}
  \bibinfo{author}{\bibfnamefont{J.~E.} \bibnamefont{Paton}},
  \bibinfo{journal}{Phys. Rev.} \textbf{\bibinfo{volume}{D31}},
  \bibinfo{pages}{2910} (\bibinfo{year}{1985}).

\bibitem[{\citenamefont{Barnes et~al.}(1995)\citenamefont{Barnes, Close, and
  Swanson}}]{Barnes:1995hc}
\bibinfo{author}{\bibfnamefont{T.}~\bibnamefont{Barnes}},
  \bibinfo{author}{\bibfnamefont{F.}~\bibnamefont{Close}}, \bibnamefont{and}
  \bibinfo{author}{\bibfnamefont{E.}~\bibnamefont{Swanson}},
  \bibinfo{journal}{Phys. Rev.} \textbf{\bibinfo{volume}{D52}},
  \bibinfo{pages}{5242} (\bibinfo{year}{1995}).

\bibitem[{\citenamefont{Horn and Mandula}(1978)}]{Horn:1977rq}
\bibinfo{author}{\bibfnamefont{D.}~\bibnamefont{Horn}} \bibnamefont{and}
  \bibinfo{author}{\bibfnamefont{J.}~\bibnamefont{Mandula}},
  \bibinfo{journal}{Phys. Rev.} \textbf{\bibinfo{volume}{D17}},
  \bibinfo{pages}{898} (\bibinfo{year}{1978}).

\bibitem[{\citenamefont{Guo et~al.}(2008)\citenamefont{Guo, Szczepaniak,
  Galata, Vassallo, and Santopinto}}]{Guo:2008yz}
\bibinfo{author}{\bibfnamefont{P.}~\bibnamefont{Guo}},
  \bibinfo{author}{\bibfnamefont{A.~P.} \bibnamefont{Szczepaniak}},
  \bibinfo{author}{\bibfnamefont{G.}~\bibnamefont{Galata}},
  \bibinfo{author}{\bibfnamefont{A.}~\bibnamefont{Vassallo}}, \bibnamefont{and}
  \bibinfo{author}{\bibfnamefont{E.}~\bibnamefont{Santopinto}},
  \bibinfo{journal}{Phys. Rev.} \textbf{\bibinfo{volume}{D78}},
  \bibinfo{pages}{056003} (\bibinfo{year}{2008}).

\bibitem[{\citenamefont{Tye}(1976)}]{Tye:1975fz}
\bibinfo{author}{\bibfnamefont{S.}~\bibnamefont{Tye}}, \bibinfo{journal}{Phys.
  Rev.} \textbf{\bibinfo{volume}{D13}}, \bibinfo{pages}{3416}
  (\bibinfo{year}{1976}).

\bibitem[{\citenamefont{Giles and Tye}(1977)}]{Giles:1977mp}
\bibinfo{author}{\bibfnamefont{R.}~\bibnamefont{Giles}} \bibnamefont{and}
  \bibinfo{author}{\bibfnamefont{S.}~\bibnamefont{Tye}},
  \bibinfo{journal}{Phys. Rev.} \textbf{\bibinfo{volume}{D16}},
  \bibinfo{pages}{1079} (\bibinfo{year}{1977}).

\bibitem[{\citenamefont{Buchmuller and Tye}(1980)}]{Buchmuller:1979gy}
\bibinfo{author}{\bibfnamefont{W.}~\bibnamefont{Buchmuller}} \bibnamefont{and}
  \bibinfo{author}{\bibfnamefont{S.}~\bibnamefont{Tye}},
  \bibinfo{journal}{Phys. Rev. Lett.} \textbf{\bibinfo{volume}{44}},
  \bibinfo{pages}{850} (\bibinfo{year}{1980}).

\bibitem[{\citenamefont{Buchmuller and Tye}(1981)}]{Buchmuller:1980su}
\bibinfo{author}{\bibfnamefont{W.}~\bibnamefont{Buchmuller}} \bibnamefont{and}
  \bibinfo{author}{\bibfnamefont{S.~H.~H.} \bibnamefont{Tye}},
  \bibinfo{journal}{Phys. Rev.} \textbf{\bibinfo{volume}{D24}},
  \bibinfo{pages}{132} (\bibinfo{year}{1981}).

\bibitem[{\citenamefont{Kalashnikova and Nefediev}(2008)}]{Kalashnikova:2008qr}
\bibinfo{author}{\bibfnamefont{Y.}~\bibnamefont{Kalashnikova}}
  \bibnamefont{and} \bibinfo{author}{\bibfnamefont{A.}~\bibnamefont{Nefediev}},
  \bibinfo{journal}{Phys. Rev.} \textbf{\bibinfo{volume}{D77}},
  \bibinfo{pages}{054025} (\bibinfo{year}{2008}).

\bibitem[{\citenamefont{Brown and Cahn}(1975)}]{Brown:1975dz}
\bibinfo{author}{\bibfnamefont{L.~S.} \bibnamefont{Brown}} \bibnamefont{and}
  \bibinfo{author}{\bibfnamefont{R.~N.} \bibnamefont{Cahn}},
  \bibinfo{journal}{Phys. Rev. Lett.} \textbf{\bibinfo{volume}{35}},
  \bibinfo{pages}{1} (\bibinfo{year}{1975}).

\bibitem[{\citenamefont{Vijande et~al.}(2005)\citenamefont{Vijande, Fernandez,
  and Valcarce}}]{Vijande:2004he}
\bibinfo{author}{\bibfnamefont{J.}~\bibnamefont{Vijande}},
  \bibinfo{author}{\bibfnamefont{F.}~\bibnamefont{Fernandez}},
  \bibnamefont{and} \bibinfo{author}{\bibfnamefont{A.}~\bibnamefont{Valcarce}},
  \bibinfo{journal}{J. Phys.} \textbf{\bibinfo{volume}{G31}},
  \bibinfo{pages}{481} (\bibinfo{year}{2005}).

\bibitem[{\citenamefont{Valcarce et~al.}(2005)\citenamefont{Valcarce,
  Garcilazo, Fernandez, and Gonzalez}}]{Valcarce:2005em}
\bibinfo{author}{\bibfnamefont{A.}~\bibnamefont{Valcarce}},
  \bibinfo{author}{\bibfnamefont{H.}~\bibnamefont{Garcilazo}},
  \bibinfo{author}{\bibfnamefont{F.}~\bibnamefont{Fernandez}},
  \bibnamefont{and} \bibinfo{author}{\bibfnamefont{P.}~\bibnamefont{Gonzalez}},
  \bibinfo{journal}{Rept. Prog. Phys.} \textbf{\bibinfo{volume}{68}},
  \bibinfo{pages}{965} (\bibinfo{year}{2005}).

\bibitem[{\citenamefont{Segovia
  et~al.}(2013{\natexlab{a}})\citenamefont{Segovia, R.~Entem, Fernandez, and
  Hernandez}}]{Segovia:2013wma}
\bibinfo{author}{\bibfnamefont{J.}~\bibnamefont{Segovia}},
  \bibinfo{author}{\bibfnamefont{D.}~\bibnamefont{R.~Entem}},
  \bibinfo{author}{\bibfnamefont{F.}~\bibnamefont{Fernandez}},
  \bibnamefont{and}
  \bibinfo{author}{\bibfnamefont{E.}~\bibnamefont{Hernandez}},
  \bibinfo{journal}{Int. J. Mod. Phys.} \textbf{\bibinfo{volume}{E22}},
  \bibinfo{pages}{1330026} (\bibinfo{year}{2013}{\natexlab{a}}).

\bibitem[{\citenamefont{Fernandez et~al.}(1992)\citenamefont{Fernandez,
  Valcarce, Gonzalez, and Vento}}]{Fernandez:1992xs}
\bibinfo{author}{\bibfnamefont{F.}~\bibnamefont{Fernandez}},
  \bibinfo{author}{\bibfnamefont{A.}~\bibnamefont{Valcarce}},
  \bibinfo{author}{\bibfnamefont{P.}~\bibnamefont{Gonzalez}}, \bibnamefont{and}
  \bibinfo{author}{\bibfnamefont{V.}~\bibnamefont{Vento}},
  \bibinfo{journal}{Phys. Lett.} \textbf{\bibinfo{volume}{B287}},
  \bibinfo{pages}{35} (\bibinfo{year}{1992}).

\bibitem[{\citenamefont{Garcilazo et~al.}(2001)\citenamefont{Garcilazo,
  Valcarce, and Fernandez}}]{Garcilazo:2001md}
\bibinfo{author}{\bibfnamefont{H.}~\bibnamefont{Garcilazo}},
  \bibinfo{author}{\bibfnamefont{A.}~\bibnamefont{Valcarce}}, \bibnamefont{and}
  \bibinfo{author}{\bibfnamefont{F.}~\bibnamefont{Fernandez}},
  \bibinfo{journal}{Phys. Rev.} \textbf{\bibinfo{volume}{C63}},
  \bibinfo{pages}{035207} (\bibinfo{year}{2001}), \bibinfo{note}{{\it ibid.}
  Phys. Rev. C64, 058201 (2001).}

\bibitem[{\citenamefont{Vijande et~al.}(2004)\citenamefont{Vijande, Garcilazo,
  Valcarce, and Fernandez}}]{Vijande:2004at}
\bibinfo{author}{\bibfnamefont{J.}~\bibnamefont{Vijande}},
  \bibinfo{author}{\bibfnamefont{H.}~\bibnamefont{Garcilazo}},
  \bibinfo{author}{\bibfnamefont{A.}~\bibnamefont{Valcarce}}, \bibnamefont{and}
  \bibinfo{author}{\bibfnamefont{F.}~\bibnamefont{Fernandez}},
  \bibinfo{journal}{Phys. Rev.} \textbf{\bibinfo{volume}{D70}},
  \bibinfo{pages}{054022} (\bibinfo{year}{2004}).

\bibitem[{\citenamefont{Segovia
  et~al.}(2008{\natexlab{a}})\citenamefont{Segovia, Yasser, R.~Entem, and
  Fernandez}}]{Segovia:2008zz}
\bibinfo{author}{\bibfnamefont{J.}~\bibnamefont{Segovia}},
  \bibinfo{author}{\bibfnamefont{A.}~\bibnamefont{Yasser}},
  \bibinfo{author}{\bibfnamefont{D.}~\bibnamefont{R.~Entem}}, \bibnamefont{and}
  \bibinfo{author}{\bibfnamefont{F.}~\bibnamefont{Fernandez}},
  \bibinfo{journal}{Phys. Rev.} \textbf{\bibinfo{volume}{D78}},
  \bibinfo{pages}{114033} (\bibinfo{year}{2008}{\natexlab{a}}).

\bibitem[{\citenamefont{Segovia et~al.}(2009)\citenamefont{Segovia, Yasser,
  R.~Entem, and Fernandez}}]{Segovia:2009zz}
\bibinfo{author}{\bibfnamefont{J.}~\bibnamefont{Segovia}},
  \bibinfo{author}{\bibfnamefont{A.}~\bibnamefont{Yasser}},
  \bibinfo{author}{\bibfnamefont{D.}~\bibnamefont{R.~Entem}}, \bibnamefont{and}
  \bibinfo{author}{\bibfnamefont{F.}~\bibnamefont{Fernandez}},
  \bibinfo{journal}{Phys. Rev.} \textbf{\bibinfo{volume}{D80}},
  \bibinfo{pages}{054017} (\bibinfo{year}{2009}).

\bibitem[{\citenamefont{Segovia
  et~al.}(2015{\natexlab{b}})\citenamefont{Segovia, R.~Entem, and
  Fernandez}}]{Segovia:2015dia}
\bibinfo{author}{\bibfnamefont{J.}~\bibnamefont{Segovia}},
  \bibinfo{author}{\bibfnamefont{D.}~\bibnamefont{R.~Entem}}, \bibnamefont{and}
  \bibinfo{author}{\bibfnamefont{F.}~\bibnamefont{Fernandez}},
  \bibinfo{journal}{Phys. Rev.} \textbf{\bibinfo{volume}{D91}},
  \bibinfo{pages}{094020} (\bibinfo{year}{2015}{\natexlab{b}}).

\bibitem[{\citenamefont{Segovia
  et~al.}(2011{\natexlab{a}})\citenamefont{Segovia, R.~Entem, and
  Fernandez}}]{Segovia:2011zza}
\bibinfo{author}{\bibfnamefont{J.}~\bibnamefont{Segovia}},
  \bibinfo{author}{\bibfnamefont{D.}~\bibnamefont{R.~Entem}}, \bibnamefont{and}
  \bibinfo{author}{\bibfnamefont{F.}~\bibnamefont{Fernandez}},
  \bibinfo{journal}{Phys. Rev.} \textbf{\bibinfo{volume}{D83}},
  \bibinfo{pages}{114018} (\bibinfo{year}{2011}{\natexlab{a}}).

\bibitem[{\citenamefont{Segovia
  et~al.}(2012{\natexlab{a}})\citenamefont{Segovia, R.~Entem, and
  Fernandez}}]{Segovia:2012cd}
\bibinfo{author}{\bibfnamefont{J.}~\bibnamefont{Segovia}},
  \bibinfo{author}{\bibfnamefont{D.}~\bibnamefont{R.~Entem}}, \bibnamefont{and}
  \bibinfo{author}{\bibfnamefont{F.}~\bibnamefont{Fernandez}},
  \bibinfo{journal}{Phys. Lett.} \textbf{\bibinfo{volume}{B715}},
  \bibinfo{pages}{322} (\bibinfo{year}{2012}{\natexlab{a}}).

\bibitem[{\citenamefont{Segovia
  et~al.}(2013{\natexlab{b}})\citenamefont{Segovia, R.~Entem, and
  Fernandez}}]{Segovia:2013kg}
\bibinfo{author}{\bibfnamefont{J.}~\bibnamefont{Segovia}},
  \bibinfo{author}{\bibfnamefont{D.}~\bibnamefont{R.~Entem}}, \bibnamefont{and}
  \bibinfo{author}{\bibfnamefont{F.}~\bibnamefont{Fernandez}},
  \bibinfo{journal}{Nucl. Phys.} \textbf{\bibinfo{volume}{A915}},
  \bibinfo{pages}{125} (\bibinfo{year}{2013}{\natexlab{b}}).

\bibitem[{\citenamefont{Segovia
  et~al.}(2011{\natexlab{b}})\citenamefont{Segovia, Albertus, R.~Entem,
  Fernandez, Hernandez et~al.}}]{Segovia:2011dg}
\bibinfo{author}{\bibfnamefont{J.}~\bibnamefont{Segovia}},
  \bibinfo{author}{\bibfnamefont{C.}~\bibnamefont{Albertus}},
  \bibinfo{author}{\bibfnamefont{D.}~\bibnamefont{R.~Entem}},
  \bibinfo{author}{\bibfnamefont{F.}~\bibnamefont{Fernandez}},
  \bibinfo{author}{\bibfnamefont{E.}~\bibnamefont{Hernandez}},
  \bibnamefont{et~al.}, \bibinfo{journal}{Phys. Rev.}
  \textbf{\bibinfo{volume}{D84}}, \bibinfo{pages}{094029}
  (\bibinfo{year}{2011}{\natexlab{b}}).

\bibitem[{\citenamefont{Segovia
  et~al.}(2012{\natexlab{b}})\citenamefont{Segovia, Albertus, Hernandez,
  Fernandez, and R.~Entem}}]{Segovia:2012yh}
\bibinfo{author}{\bibfnamefont{J.}~\bibnamefont{Segovia}},
  \bibinfo{author}{\bibfnamefont{C.}~\bibnamefont{Albertus}},
  \bibinfo{author}{\bibfnamefont{E.}~\bibnamefont{Hernandez}},
  \bibinfo{author}{\bibfnamefont{F.}~\bibnamefont{Fernandez}},
  \bibnamefont{and} \bibinfo{author}{\bibfnamefont{D.}~\bibnamefont{R.~Entem}},
  \bibinfo{journal}{Phys. Rev.} \textbf{\bibinfo{volume}{D86}},
  \bibinfo{pages}{014010} (\bibinfo{year}{2012}{\natexlab{b}}).

\bibitem[{\citenamefont{Segovia
  et~al.}(2013{\natexlab{c}})\citenamefont{Segovia, Hernandez, Fernandez, and
  R.~Entem}}]{Segovia:2013sxa}
\bibinfo{author}{\bibfnamefont{J.}~\bibnamefont{Segovia}},
  \bibinfo{author}{\bibfnamefont{E.}~\bibnamefont{Hernandez}},
  \bibinfo{author}{\bibfnamefont{F.}~\bibnamefont{Fernandez}},
  \bibnamefont{and} \bibinfo{author}{\bibfnamefont{D.}~\bibnamefont{R.~Entem}},
  \bibinfo{journal}{Phys. Rev.} \textbf{\bibinfo{volume}{D87}},
  \bibinfo{pages}{114009} (\bibinfo{year}{2013}{\natexlab{c}}).

\bibitem[{\citenamefont{Segovia
  et~al.}(2008{\natexlab{b}})\citenamefont{Segovia, Entem, and
  Fernandez}}]{Segovia:2008zza}
\bibinfo{author}{\bibfnamefont{J.}~\bibnamefont{Segovia}},
  \bibinfo{author}{\bibfnamefont{D.}~\bibnamefont{Entem}}, \bibnamefont{and}
  \bibinfo{author}{\bibfnamefont{F.}~\bibnamefont{Fernandez}},
  \bibinfo{journal}{Phys. Lett.} \textbf{\bibinfo{volume}{B662}},
  \bibinfo{pages}{33} (\bibinfo{year}{2008}{\natexlab{b}}).

\bibitem[{\citenamefont{Bali et~al.}(2005)\citenamefont{Bali, Neff, Duessel,
  Lippert, and Schilling}}]{Bali:2005fu}
\bibinfo{author}{\bibfnamefont{G.~S.} \bibnamefont{Bali}},
  \bibinfo{author}{\bibfnamefont{H.}~\bibnamefont{Neff}},
  \bibinfo{author}{\bibfnamefont{T.}~\bibnamefont{Duessel}},
  \bibinfo{author}{\bibfnamefont{T.}~\bibnamefont{Lippert}}, \bibnamefont{and}
  \bibinfo{author}{\bibfnamefont{K.}~\bibnamefont{Schilling}}
  (\bibinfo{collaboration}{SESAM}), \bibinfo{journal}{Phys. Rev.}
  \textbf{\bibinfo{volume}{D71}}, \bibinfo{pages}{114513}
  (\bibinfo{year}{2005}).

\bibitem[{\citenamefont{Eichten et~al.}(1978)\citenamefont{Eichten, Gottfried,
  Kinoshita, Lane, and Yan}}]{Eichten:1978tg}
\bibinfo{author}{\bibfnamefont{E.}~\bibnamefont{Eichten}},
  \bibinfo{author}{\bibfnamefont{K.}~\bibnamefont{Gottfried}},
  \bibinfo{author}{\bibfnamefont{T.}~\bibnamefont{Kinoshita}},
  \bibinfo{author}{\bibfnamefont{K.}~\bibnamefont{Lane}}, \bibnamefont{and}
  \bibinfo{author}{\bibfnamefont{T.-M.} \bibnamefont{Yan}},
  \bibinfo{journal}{Phys. Rev.} \textbf{\bibinfo{volume}{D17}},
  \bibinfo{pages}{3090} (\bibinfo{year}{1978}).

\bibitem[{\citenamefont{Eichten et~al.}(1980)\citenamefont{Eichten, Gottfried,
  Kinoshita, Lane, and Yan}}]{Eichten:1979ms}
\bibinfo{author}{\bibfnamefont{E.}~\bibnamefont{Eichten}},
  \bibinfo{author}{\bibfnamefont{K.}~\bibnamefont{Gottfried}},
  \bibinfo{author}{\bibfnamefont{T.}~\bibnamefont{Kinoshita}},
  \bibinfo{author}{\bibfnamefont{K.}~\bibnamefont{Lane}}, \bibnamefont{and}
  \bibinfo{author}{\bibfnamefont{T.-M.} \bibnamefont{Yan}},
  \bibinfo{journal}{Phys. Rev.} \textbf{\bibinfo{volume}{D21}},
  \bibinfo{pages}{203} (\bibinfo{year}{1980}).

\bibitem[{\citenamefont{Aidala et~al.}(2011)\citenamefont{Aidala, Ellinghaus,
  Sassot, Seele, and Stratmann}}]{Aidala:2010bn}
\bibinfo{author}{\bibfnamefont{C.~A.} \bibnamefont{Aidala}},
  \bibinfo{author}{\bibfnamefont{F.}~\bibnamefont{Ellinghaus}},
  \bibinfo{author}{\bibfnamefont{R.}~\bibnamefont{Sassot}},
  \bibinfo{author}{\bibfnamefont{J.~P.} \bibnamefont{Seele}}, \bibnamefont{and}
  \bibinfo{author}{\bibfnamefont{M.}~\bibnamefont{Stratmann}},
  \bibinfo{journal}{Phys. Rev.} \textbf{\bibinfo{volume}{D83}},
  \bibinfo{pages}{034002} (\bibinfo{year}{2011}).

\end{thebibliography}

\end{document}